# Low-Frequency Electronic Noise Spectroscopy of Quasi-2D van der Waals Antiferromagnetic Semiconductors

Subhajit Ghosh[1], Fariborz Kargar[1], Amirmahdi Mohammadzadeh[1], Sergey Rumyantsev[2] and Alexander A. Balandin[1,3,♦]

[1]Nano-Device Laboratory (NDL), Department of Electrical and Computer Engineering, Bourns College of Engineering, University of California, Riverside, California 92521 USA

[2]CENTERA Laboratories, Institute of High-Pressure Physics, Polish Academy of Sciences, Warsaw 01-142 Poland

[3]Phonon Optimized Engineered Materials (POEM) Center, Materials Science and Engineering Program, University of California – Riverside, California 92521 USA

## Abstract

We investigated low-frequency current fluctuations, *i.e.* noise, in the quasi-two-dimensional (2D) van der Waals antiferromagnetic semiconductor $FePS_3$ with the electronic bandgap of 1.5 eV. The electrical and noise characteristics of the *p*-type, highly resistive, thin films of $FePS_3$ were measured at different temperatures. The noise spectral density was of the $1/f$-type over most of the examined temperature range but revealed well-defined Lorentzian bulges, and increased strongly near the Néel temperature $T_N$=118 K ($f$ is the frequency). Intriguingly, the noise spectral density attained its minimum at temperature $T$~200 K, which was attributed to an interplay of two opposite trends in noise scaling – one for semiconductors and another for materials with the phase transitions. The Lorentzian corner frequencies revealed unusual dependence on temperature and bias voltage, suggesting that their origin is different from the generation – recombination noise in conventional semiconductors. The obtained results are important for proposed applications of antiferromagnetic semiconductors in spintronic devices. They also attest to the power of the *noise spectroscopy* for monitoring various phase transitions.

---

♦ Corresponding author (AAB): balandin@ece.ucr.edu ; http://balandingroup.ucr.edu

Transition-metal phospho-trichalcogenides, $MPX_3$, where M is a transition metal *e.g.* V, Mn, Fe, Co, Ni, or Zn and X is a chalcogenide as S, Se, Te, have recently attracted a lot of attention [1-6]. These layered quasi-two-dimensional (2D) van der Waals (vdW) compounds have interesting electronic, optical, and magnetic properties that can offer new device functionalities [7-19]. It has been demonstrated that some $MPX_3$ thin films are one of the rare few-layer vdW materials, which can have stable intrinsic antiferomagnetism (AFM) even at mono- and few layer thicknesses [20-22]. The existence of weak vdW bonds between the $MPX_3$ layers makes them potential candidates for the 2D spintronic devices. The metal element of the $MPX_3$ materials modifies the band gap from a medium band gap of ~1.3 eV to a wide band gap of ~3.5 eV [1-3]. The diverse properties of these materials tunable by proper selection and combination of the M and X elements make the $MPX_3$ materials an interesting platform for fundamental science and practical applications [1-4, 23-30].

Among $MPX_3$ materials, $FePS_3$ is particularly promising. Its Ising-type ordering allows it to maintain the bulk-like magnetic behavior down to a single monolayer, which explains its moniker - "magnetic graphene" [31]. The cleavage energy of $FePSe_3$ is slightly higher than that of graphite, while that for all other combinations of the M and X elements is lower than that of graphite [32]. The Néel temperature, $T_N$, for $FePS_3$ is reported to be around 118 K [1,2,20]. It shows strong magnetic anisotropy [15,33,34]. In the crystal, each Fe atom is ferromagnetically coupled with two of its neighbors but the layer is antiferromagnetically coupled with nearest layers making it a zigzag-AFM or z-AFM material [1]. The semiconducting nature of $FePS_3$, with the energy band gap of 1.5 eV, and a possibility of the strain engineering of electron and phonon band-structure create additional means for the control of both electron and spin transport [9,17,35]. The quantized spin waves, *i.e.*, magnons in $FePS_3$, have frequencies in the terahertz (THz) frequency range [36,37], which makes this material interesting for THz magnonic devices. However, the electron and spin transport properties of $FePS_3$ have not been studied in sufficient details yet to assess the potential of such materials for THz applications.

Here, we report the results of investigation of low-frequency current fluctuations, *i.e.* electronic noise, in thin films of $FePS_3$. The I-Vs and noise characteristics were measured as a function of temperature to understand their evolution below the room temperature (RT) and, particularly in the vicinity of the Néel temperature of the transition from paramagnetic (PM) to AFM ordering. The knowledge of the low-frequency noise is important for assessing the quality of the material and its prospects for any electronic, spintronic or sensor application [38,39]. The noise spectral density and its dependence on external stimuli, *e.g.* electrical bias and temperature, can shed light on electron transport, carrier recombination mechanisms and, what is most important in this case, magnetic and metal-insulator phase transitions. We have previously used successfully the low-frequency noise measurements as the "noise spectroscopy" for monitoring phase transitions in the 2D charge-density-wave materials [40-42]; examining the specifics of magnon transport in magnetic electrical insulators [43]; and clarifying the nature of electron transport in quasi-one-dimensional (1D) vdW materials [44,45]. Our measurements of noise in thin films of $FePS_3$ reveal a number of interesting features, which contribute to a better understanding of the properties of this AFM vdW semiconductor.

For this study, we used high-quality single crystals of $FePS_3$ (HQ Graphene), which were synthesized by the chemical vapor deposition (CVD). The crystals had background *p*-type doping. Figure 1 (a) shows the schematic of the monoclinic atomic crystal structure of $FePS_3$ with C2/m symmetry. The Fe, P, and S atoms are indicated by the violet, green and yellow circles, respectively. The red and violet arrows indicate the up and down directions of the z-AFM spin ordering [1,20]. The X-ray diffraction (XRD) and Raman spectroscopy were used to verify the material composition and crystal structure (see Figures 1 (b) and (c)). The Raman spectra were taken in the backscattering configuration under 488 nm laser excitation. It is known that $FePS_3$ crystal contains 30 irreducible zone center phonons at Γ point of the first Brillouin zone: $\Gamma = 8A_g + 6A_u + 7B_g + 9B_u$, among which only the $A_g$ and $B_g$ modes are Raman active [46,47]. The Raman active modes can be classified into the high-frequency phonon modes, which belong to the internal vibrations of the $(P_2S_6)^{4-}$ anions, and the low-frequency phonon modes, typically below 200 cm$^{-1}$, which correspond to the interaction of the transition metals (Fe) with both the P and S atoms In our measured Raman spectrum, we were able to observe seven Raman frequency peaks: 98 cm$^{-1}$ ($A_g$, $B_g$), 158 cm$^{-1}$ ($A_g$, $B_g$), 226 cm$^{-1}$ ($A_g$, $B_g$), 247 cm$^{-1}$ ($A_g$), 279 cm$^{-1}$ ($A_g$, $B_g$), 380 cm$^{-1}$ ($A_g$) and 579 cm$^{-1}$ ($A_g$). The sharp peak at 520 cm$^{-1}$ comes from the $Si/SiO_2$ substrate. The obtained Raman data are in line with the literature reports attesting to the quality of the crystals [6,20,46-49].

The test structures were prepared by mechanically exfoliating bulk crystals onto a Si/SiO$_2$ substrate with the oxide thickness of 300 nm. The uniform and relatively thick layers (200 nm – 400 nm) of the exfoliated layers were chosen in order to compensate for the high resistivity of the material and increase the current level in the device channels. The electron beam lithography (EBL) was used to fabricate the device structures with four contacts. The contacts were fabricated using Cr/Au metal stacks, with the 2-µm distance between the nearest contacts. The optical microscopy image of the tested device is shown in Figure 1 (d). The thickness of the channel of this device structure was determined with the atomic force microscopy (AFM) to be ~300 nm. Two terminal I-V characteristics were measured in vacuum inside a cryogenic probe station (Lake Shore TTPX) in the temperature range from RT down to ~ 110 K.

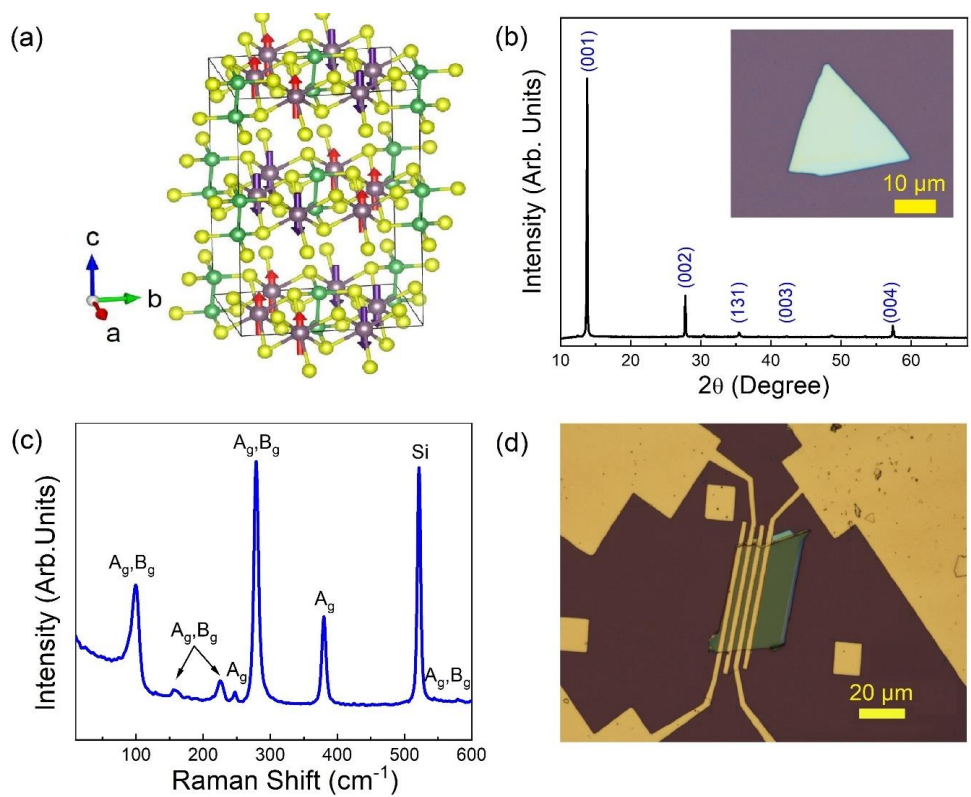

**Figure 1: Material characteristics and test structure.** (a) Crystal structure of bulk FePS$_3$ material. (b) The XRD data for the FePS$_3$ crystal. The inset shows an optical image of a representative crystal. (c) Raman spectrum of the exfoliated FePS$_3$ film on Si/SiO$_2$ substrate. The Raman data were accumulated under 488-nm laser excitation at room temperature. (d) Optical microscopy image of the fabricated test structure, containing four Cr/Au contacts.

Figure 2 (a) shows the I-V measurement plots at lower bias voltages (≤ 5 V) between 300 K and 150 K where the IV curves are nearly linear. Figure 2 (b) shows measured I-V curves at different temperatures at the extended voltage regimes, in the range from RT down to 110 K. The I-Vs in semi-logarithmic scale are presented in Figure 2 (c). The I-V characteristics are nearly linear at small bias voltage as discussed above but became non-linear for voltages above 10 V. Similar non-linear I-Vs have been reported for other $MPX_3$ materials [25-27,50,51]. The study, which reported linear I-Vs for $FePS_3$ devices, was limited to the lower voltage regimes [52]. The exact nature of the non-linearity at high bias voltage is still an open question. While some studies attributed it to the formation of the Schottky barriers at the junction between the channel and the metal contacts [25-27,50], the I-V characteristics of our devices plotted in semi-logarithmic scale (see Figure 2 (b)) suggest deviation from the conventional Schottky barrier thermionic model. Other possible reasons for non-linearity can be related to both intrinsic properties of this material as well as to such effects as surface charge accumulation under high electric bias. The local Joule heating does not seem to be a likely mechanism due to the small level of the electric currents even though $MPX_3$ materials have low thermal conductivity, which can be further reduced in thin films owing to the phonon – boundary scattering [6].

One can also see from Figure 2 (b - d) that the resistivity of the $FePS_3$ channel is the smallest at the highest temperature, and increases as the temperature is reduced. This is expected owing to the stronger thermal generation of the charge carriers at higher temperatures, which increases the concentration of free carriers contributing to the current [53]. The increase in the resistivity appears to be particularly strong as the temperature reduces below T=200 K (see Supplemental Figures S1 and S2). Similar temperature dependence of resistivity in such material has been reported previously [11,54]. For additional I-V characteristics, close to the Néel transition temperature $T_N$ = 118 K, see the Supplemental Figure S3.

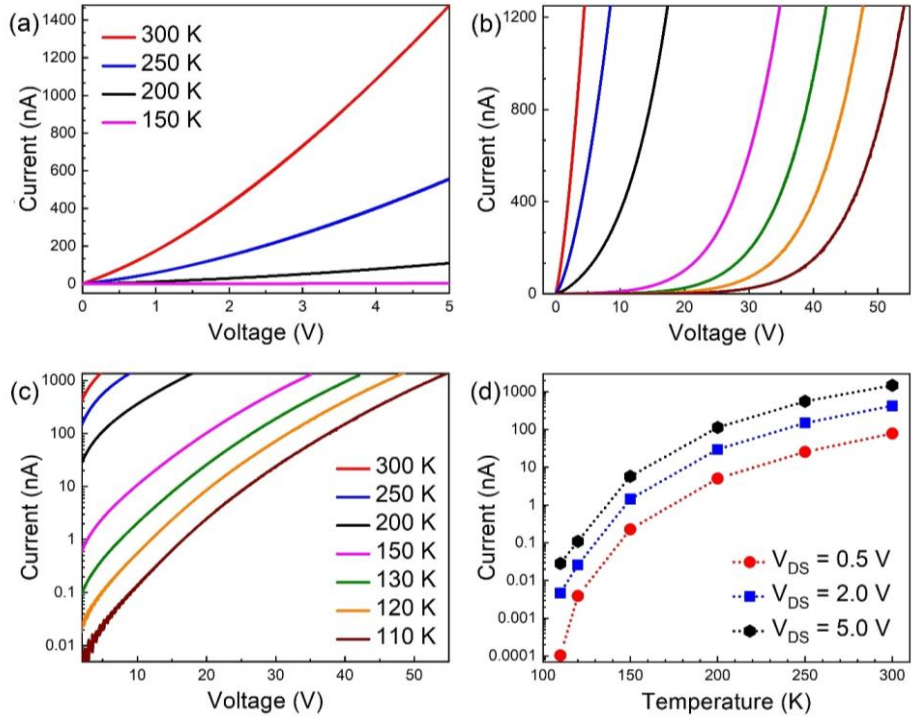

**Figure 2: Current-voltage characteristics of FePS$_3$ antiferromagnetic semiconductor.** (a) Current-voltage characteristics of the FePS$_3$ test structure measured at low voltage bias (≤5V) in the temperature range from 150 K to 300 K. The I-V plots are almost linear at low voltages. The current at 150 K is small due to the high resistance of the FePS$_3$ channel. (b) Current-voltage characteristics of the same test structure over extended bias range, measured in the temperature range from 110 K to 300 K. (c) The same characteristics as in panel (b) but shown in the semi-log scale. The temperature color legend is the same for panel (b) and (c). The non-linearity of I-V characteristics is pronounced at large biases. The resistance of the material increases with decreasing temperature. (d) The current through the FePS$_3$ channel as a function of temperature at fixed bias voltage.

The low-frequency noise measurements of the fabricated FePS$_3$ devices were carried out using an in-house experimental setup. The schematic of the setup is shown in Figure 3 (a). The equipment consists of six 12 V batteries, with the low internal noise, connected in series to supply a maximum of 72 V bias to the system; a potentiometer to control the voltage drop across the circuit; a load resistor; a low noise voltage preamplifier; and a dynamic signal analyzer. The batteries were intentionally used as the voltage source instead of connecting the system directly to the DC power source in order to minimize the effect of 60 Hz power line frequency and its harmonics. Details of our measurement protocol are described in the Methods section and in prior reports in the context of other materials systems [40-42,44,45,55-61]. Measuring low-frequency noise in conductors with high electrical resistivity is challenging – one needs to have a sufficient current level to obtain reliable data. From the other side, one

6 | P a g e

typically prefers to use the linear region of the I-V characteristics for biasing the device during the noise measurements.

The low-frequency noise measured at RT for different bias voltages, $V_D$, varying from 0.14 V to 3.73 V is shown in Figure 3 (b). The voltage-referred noise-power spectral density, $S_V$, reveals the typical flicker noise trend, $S_V \sim 1/f^\gamma$, where $f$ is the frequency and parameter $\gamma \approx 1$. The $1/f$ noise is typical for many semiconductor and metallic materials [62]. We verified that the $1/f$ noise level at the lowest applied bias, $V_D$=0.14 mV, is more than an order of magnitude higher than the background noise of the measurement system (refer to the Supplemental Figure S4), confirming that the spectra in Figure 3 (b) are indeed resulting from the current fluctuations in the device under test.

Figure 3 (c) presents the current noise-power spectral density, $S_I$, as a function of current, $I_D$, at different temperatures. The data are shown at fixed frequency $f = 10$ Hz. In general, the dependence of $S_I$ on the device current $I_D$ is expected to be quadratic, *i.e.* $S_I \sim I_D^2$, so that the slope in the $S_I$ vs $I_D$ plot is close to 2. It is observed in various materials where the noise is $1/f$ type, without the superimposed G-R bulges [42,45,55]. In our measurements the $S_I(I_D)$ dependence at different temperatures does not show a perfect quadratic scaling, Instead, we obtained $S_I(I_D) \sim I_D^\zeta$ where parameter $\zeta$ is in the range between from 1.49 to 2.15. We attribute this deviation to the non-linearity of I-V characteristics in the tested devices. It is not related to any current induced damage in the devices, *e.g.* onset of electromigration owing to the small current levels used in the measurements. The reproducibility of the noise data also attests to the absence of any damage during the measurements.

The normalized noise-power spectral densities, $S_I/I^2$, as a function of frequency at different temperatures are presented in Figure 3 (d). These spectra were measured at a fixed current through the channel, $I \equiv I_D$=50 nA. The noise measurements were intentionally carried out at the smallest current level to avoid even small Joule heating of the channel near the expected phase transition temperature points. At low temperatures, the measurements were conducted at 2-K interval in order to examine the effect of transition from PM to AFM phase at the Néel temperature of 118 K. The spectra show the $1/f$ noise as temperature decreases from 300 K to 130 K. As it approaches $T_N$=118 K, the spectrum develops pronounced Lorentzian bulges. The spectral position of the Lorentzian bulges reveals a pronounced dependence on temperature.

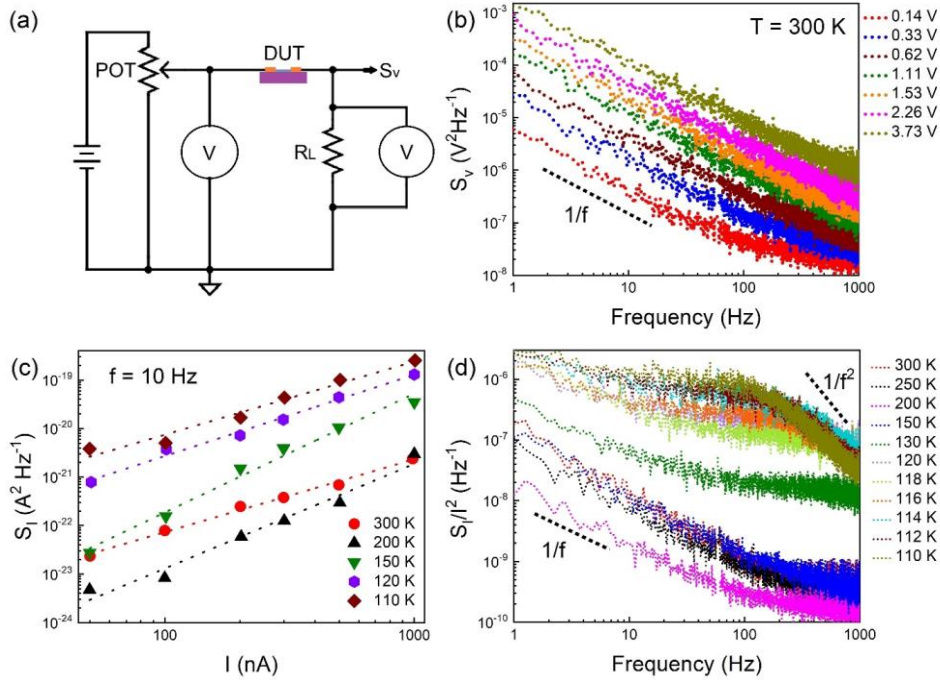

**Figure 3: Low-frequency noise spectra of FePS$_3$ thin films.** (a) Schematic of the noise measurements showing the biasing of the device under test (DUT). (b) The voltage-referred noise spectral density, $S_V$, as a function of frequency measured for different bias voltage at room temperature. (c) The current noise spectral density, $S_I$, as a function of the channel current at different temperatures and fixed frequency $f$=10 Hz. (d) The normalized current noise spectral density, $S_I/I^2$, as a function of frequency at different temperatures. The noise was measured at the constant channel current $I_D$ = 50 nA. Note the appearance of Lorentzian bulges at low temperatures.

The Lorentzian shape of the noise spectrum at $T$=120 K and below can be described by the equation $S_I(f) = S_0 \times (f_c^2/(f^2 + f_c^2))$ where $f_c$ is the corner frequency of the spectrum defined as $f_c = (2\pi\tau)^{-1}$, $S_0$ is the frequency independent portion of the function $S_I(f)$ when $f < f_c$ and $\tau$ represents the characteristic time constant [41,44]. In most cases, the emergence of the Lorentzian bulges in semiconductor devices is associated with the generation – recombination (G-R) noise, which is conventionally interpreted as evidence that the material does have high concentration of certain defects, which would dominate the noise properties [62]. However, this is not the only possible origin of the Lorentzian bulges. They can also be related to various phase transitions that material is undergoing under changing temperature or applied electrical bias [41,42,44]. We have previously observed Lorentzian spectra associated with the charge-density-wave (CDW) phase transitions [41,42], magnon transport in electrically insulating materials [43], and electron transport in bundles of quasi-1D vdW materials [44]. The noise mechanisms leading to the appearance of Lorentzian bulges, which are associated with the phase transitions, are completely different from that of the G-R trapping noise.

To further analyze the effect of temperature and applied bias on the noise spectra, we plotted the normalized noise-power spectral density, $S_I/I^2$, multiplied by the frequency, *i.e.* $f \times S_I/I^2$ *vs. f*. This procedure helps to separate the Lorentzian features from the 1/*f* background. Figure 4 (a) illustrates how the maxima, which correspond to the corner frequencies, $f_c$, are shifting with the temperature. The data are presented for $I_D$ = 50 nA. Figure 4 (b) presents $f \times S_I/I^2$ *vs. f* for the fixed temperature *T*=110 K and varied channel current, $I_D$. For each spectrum, we determine the corner frequency by numerical fitting (see Supplemental Figures S5 – S7). The corner frequency, $f_c$, at different currents, is shown as a function of 1000/*T* in Figure 4 (c). It increases strongly near the Néel temperature and follows the same trend across all currents. The corner frequency values at 118 K and 120 K for higher currents are out of the measured frequency range. From the Arrhenius plot of *ln(f_c)* vs. 1000/*T* in Figure 4 (c), we have also extracted the "noise activation" energy, $E_A$, to be 0.9 eV for this material (refer to Supplemental Figure S8). This activation energy is unrealistically large for a semiconductor with the 1.5-eV energy bandgap. The latter confirms our conclusion that the Lorentzian noise bulges are not of G-R charge carrier trapping - de-trapping origin. Since the Lorentzian bulges were observed at temperatures close to the Neel temperature we argue that these spectral features are associated not with the GR noise but with the PM – AFM phase transition. The additional evidence for this conclusion comes from the fast increase of the corner frequency with the applied bias voltage (see Figure 4 (d)). Such a strong dependence of $f_c$ on $V_D$ is unusual for conventional trapping G-R noise. However, it is common in CDW materials where Lorentzian bulges appear at the nearly commensurate to incommensurate CDW phase transition [40-42].

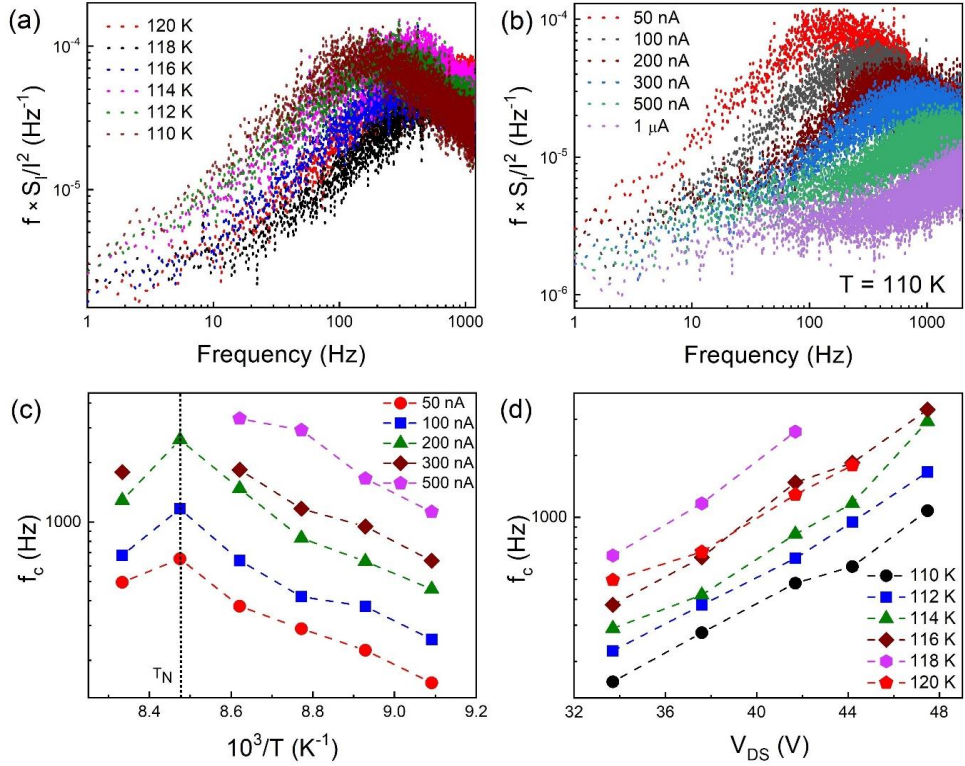

**Figure 4: Lorentzian spectral features in the noise spectra.** (a) The normalized noise spectral density multiplied by the frequency, $S_I/I^2 \times f$, as a function of frequency measured near the Néel temperature at the constant channel current $I_D$=50 nA. (b) The normalized noise spectral density multiplied by the frequency, $S_I/I^2 \times f$, as a function of frequency measured for different channel currents at a fixed temperature $T$=110 K. (c) The corner frequency, $f_c$, of the Lorentzian noise components as function of the inverse temperature, 1000/$T$, measured at different channel currents. The corner frequency attains its highest value at the Néel temperature $T_N$=118 K. The dashed lines are guides for the eye. (d) The corner frequency, $f_c$, of the Lorentzian noise components as function of the applied bias voltage measured at different temperatures. Note the strong dependence of the corner frequency on temperature and bias voltage.

Figures 5 (a) and (b) show the current fluctuations, *i.e.* noise, in the time-domain near the Néel temperature and high temperature, respectively. The large sharp peaks in both figures are due to the external electromagnetic influence since they were also observed without the electrical bias. These features should be ignored in the analysis. As one can see in Figure 5 (a), the noise has the form of the pulses with nearly constant amplitudes and random duration and intervals between the pulses. The amplitude of the pulses depends only weakly on temperature. This type of current fluctuations is referred to as the random telegraph signal (RTS) noise. The RTS noise appears when only one fluctuator within the whole sample is responsible for noise [43,63,64]. Since the assumption that the macroscopic sample includes only one charge carrier trap, which contributes to noise, is not realistic, the existence of the RTS noise is a further proof that this noise, and the corresponding Lorentzian bulges in the frequency domain, are associated with the phase transition. The RTS noise disappears at higher temperatures (Figure 5 (b)), which is in line with our explanation. It is important to note that the phase transition happens either in some local area only or within the whole sample at once. We believe that the second assumption is more realistic, *i.e.* the observed Lorentzian features are accurate metrics of the phase transition in 2D AFM semiconductors. One can envision the use of the low-frequency noise measurements for the noise spectroscopy of various phase transitions in different types of materials. In this regard, the noise can be interpreted as a signal similar to the fluctuation-enhanced sensing [65].

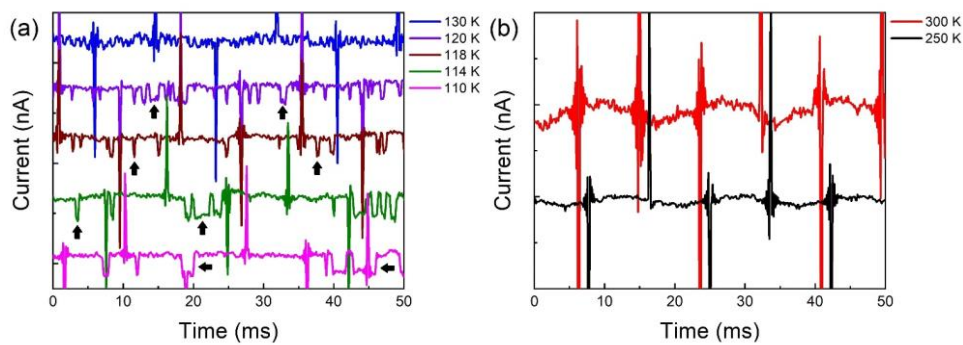

**Figure 5: Current fluctuations in the time domain.** (a) The time-domain noise spectra near the Néel temperature. The current fluctuations measured in the temperature range from 110 K to 120 K reveal clear signatures of the random telegraph signal noise. The black arrows indicate representative RTS pulses at each temperature. (b) The time-domain noise spectra near the room temperature. No random telegraph signal noise is observed. Note that in both panels, the large spikes are due to the electromagnetic interference and should be ignored in the data analysis.

We now look closely at the evolution of the normalized noise spectral density, $S_I/I^2$, as a function of temperature (see Figure 6 (a-d)). One can see a consistent trend at every frequency and current through the device channel. At $T < 200$ K, the noise increases as the temperature decreases and reaches the Néel transition temperature to AFM phase. One can expect that the noise will be close to its maximum value near the phase transition temperature, in this case, from PM to AFM phase. At this point, the material system is undergoing structural change, often characterized by an increased disorder and, as a result, increased electronic noise. The increased noise can also be associated with abrupt changes in the resistance and instability of the characteristics of the material at the phase transition. In general, it is possible that a maximum in the noise should signal a phase transition [40-42,66-70]. Prior studies indicate that the noise increases in the vicinity of the metal-insulator transition [66], spin glass transition [67,68], and other various phase transitions [69,70].

The noise temperature dependence at around $T$=200 K, where it attains minimum, is intriguing (see Figure 6 (a-d) and Supplemental Figure S9). Previously, this temperature was identified as a possible Mott metal – insulator transition point for FePS$_3$ [11,54]. The latter was established *via* measurements that involved an application of high pressure [11,54]. One would expect a noise increase at the metal – insulator transition. To explain the noise minimum near 200 K, we consider the following physical mechanism. The conventional models of the low-frequency noise in semiconductors predict the noise magnitude proportional to temperature. For example, in McWhorter's theory, the noise spectral density scales with temperature as $S_I \sim k_B T$ (here $k_B$ is the Boltzmann constant) [71]. This explains the decrease of the noise with the temperature decrease from 300 K to 200 K. The increase in the noise level with the further temperature decrease is associated with the early effect of the phase transition at Néel temperature. In this scenario, the position of the noise minimum is an interplay of two trends – one is the noise temperature dependence typical for semiconductors and the other one is the noise increase near the phase transition point. There have been reports of a similar trend in electrical noise from the spin fluctuations *vs.* temperature in CuMn [67] and Cr [68]. While noise scaling in metals is more complicated one can envision a similar interplay resulting in non-monotonic dependence of noise on temperature.

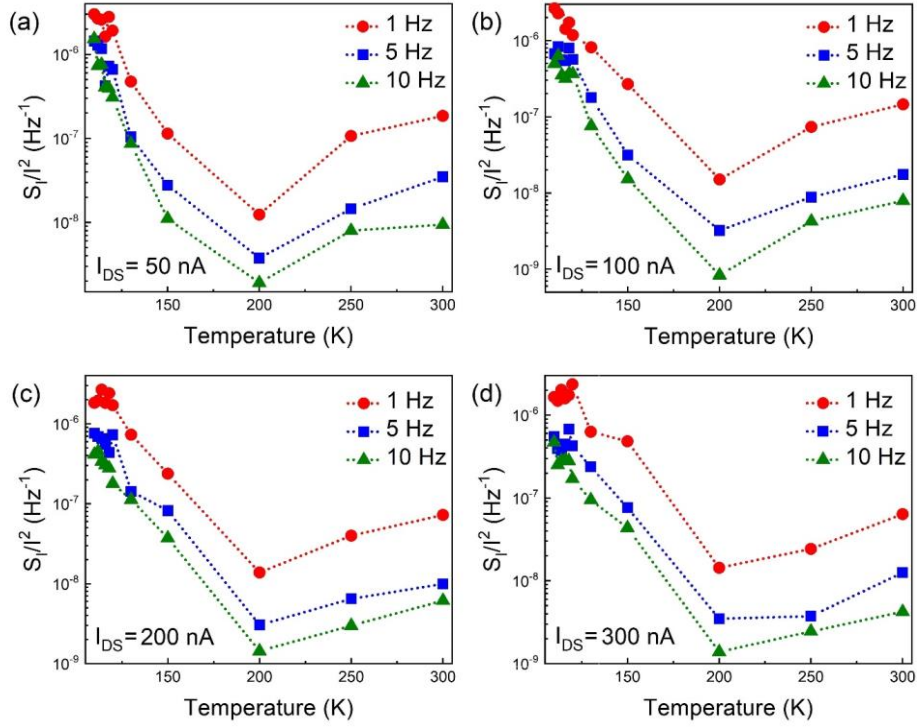

**Figure 6: Noise evolution with temperature.** The normalized current noise spectral density as a function of temperature plotted for three different frequencies $f$=1 Hz, 5 Hz and 10 Hz. The data are presented for several fixed current levels, $I_D$, to demonstrate the consistency and reproducibility: (a) 50 nA, (b) 100 nA, (c) 200 nA and (d) 300 nA. The noise level attains its maximum near the Néel temperature. The non-monotonic dependence of noise spectral density on temperature is likely related to an interplay of two opposite trends in noise scaling – one for semiconductors and another for materials with the phase transitions.

In conclusion, we investigated low-frequency noise in the quasi-2D van der Waals AFM semiconductor FePS$_3$ with the electronic bandgap of 1.5 eV. The noise spectral density was of the 1/$f$-type over most of the examined temperature range but reveals well-defined Lorentzian bulges, and it increases strongly near the Néel temperature $T_N$ =118 K. Intriguingly, the noise spectral density attained its minimum at temperature $T$~200 K, which is likely related to an interplay of two opposite trends in the noise scaling – one for conventional semiconductors and another for materials with the phase transitions. The low-frequency noise in semiconductors typically scales with temperature as $S_I$~$k_B T$. This explains the noise reduction with the temperature decrease from 300 K to 200 K. The trend reverses as the temperature continues to decrease and approaches the AFM phase transition point. The noise spectra revealed well-defined Lorentzian bulges near the Néel temperature $T_N$=118 K. The Lorentzian corner frequencies depend strongly on temperature and bias voltage, suggesting that their origin is different from the conventional generation – recombination noise. The observed Lorentzian spectral features are signatures of the AFM phase transitions rather than electron trapping and de-trapping by the defects. The obtained results are important for proposed applications of antiferromagnetic semiconductors in spintronic devices. They also attest to the power of the *noise spectroscopy* for monitoring phase transitions of various nature and in different materials.

## METHODS

**Device Fabrication:** We employed the electron beam lithography (EBL) to fabricate the devices for the noise measurements. The bulk AFM semiconductor crystals were mechanically exfoliated on top of the Si/SiO$_2$ substrates. The exfoliated layers with desired thickness were selected for device fabrications. The substrates containing FePS$_3$ flakes were spin coated with layers of A4 PMMA in a photo-resist spin coat station. The resist coated layers were then placed inside an EBL system (Leo SUPRA 55 SEM/EBL). The predesigned patterns were written on the surface of the substrates. Once completed, the substrates were developed with a standard solution. After that the metals (Cr/Au) were deposited using an electron beam evaporator tool (Temescal BJD 1800) to create the contact electrodes. The thickness of the thin film of the devices were measured using the atomic force microscopy (AFM) system (Dimension 3100).

**Noise Measurements:** The noise spectra were measured with a dynamic signal analyzer (Stanford Research 785). To minimize the 60 Hz noise and its harmonics, we used a battery biasing circuit to apply voltage bias to the devices. The devices were connected to a cryogenic probe station (Lakeshore TTPX). The I-V characteristics were measured in the cryogenic probe station with a semiconductor analyzer (Agilent B1500). The noise measurements were conducted in the two-terminal device configuration. The dynamic signal analyzer measures the absolute voltage noise spectral density, $S_V$, of a parallel resistance network of a load resistor, $R_L$, and the device under test (DUT), $R_D$. The normalized current noise spectral density, $S_I/I^2$, was calculated from the equation: $S_I/I^2 = S_V \times [(R_L + R_D)/(R_L \times R_D)]^2/(I^2 \times G^2)$, where $G$ is the amplification of the low-noise amplifier.

## Acknowledgements
The work at UC Riverside was supported, in part, by the U.S. Department of Energy, Office of Basic Energy Sciences, Division of Materials Sciences and Engineering under the contract No.

DE-SC0021020. S.R. acknowledges the support from the CENTERA Laboratories in the framework of the International Research Agendas Program for the Foundation for Polish Sciences co-financed by the European Union under the European Regional Development Fund (No. MAB/2018/9).

**Author Contributions**

A.A.B. conceived the idea, coordinated the project, and contributed to experimental data analysis; S.G. fabricated and tested devices, conducted the low-frequency noise measurements, and analyzed experimental data; S.L.R. and F.K. contributed to the experimental data analysis; A.M.M assisted with the fabrication of the devices. All authors contributed to writing of the manuscript.

**Supplementary Information**

Additional details concerning device fabrication and noise measurements are available on the journal web-site for free-of-charge.

**REFERENCES**

1. F. Wang, T. A. Shifa, P. Yu, P. He, Y. Liu, F. Wang, Z. Wang, X. Zhan, X. Lou, F. Xia, J. He, *Advanced Functional Materials* **2018**, *28*, 1802151.
2. L. Wang, P. Hu, Y. Long, Z. Liu, X. He, *J. Mater. Chem. A* **2017**, *5*, 22855.
3. R. Samal, G. Sanyal, B. Chakraborty, C. S. Rout, *J. Mater. Chem. A* **2021**, *9*, 2560.
4. Y. Zhang, T. Fan, S. Yang, F. Wang, S. Yang, S. Wang, J. Su, M. Zhao, X. Hu, H. Zhang, T. Zhai, *Small Methods* **2021**, 2001068.
5. C. C. Mayorga-Martinez, Z. Sofer, D. Sedmidubský, Š. Huber, A. Y. S. Eng, M. Pumera, *ACS Applied Materials and Interfaces* **2017**, *9*, 12563.
6. F. Kargar, E. A. Coleman, S. Ghosh, J. Lee, M. J. Gomez, Y. Liu, A. S. Magana, Z. Barani, A. Mohammadzadeh, B. Debnath, R. B. Wilson, R. K. Lake, A. A. Balandin, *ACS Nano* **2020**, *14*, 2424.
7. R. Brec, D. M. Schleich, G. Ouvrard, A. Louisy, J. Rouxel, *Inorganic Chemistry* **1979**, *18*, 1814.
8. Y. V. Kuzminskii, B. M. Voronin, N. N. Redin, *Journal of Power Sources* **1995**, *55*, 133.
9. B. L. Chittari, Y. Park, D. Lee, M. Han, A. H. Macdonald, E. Hwang, J. Jung, *Physical Review B* **2016**, *94*, 184428.
10. K. Ichimura, M. Sano, *Synthetic Metals* **1991**, *45*, 203.
11. M. Tsurubayashi, K. Kodama, M. Kano, K. Ishigaki, Y. Uwatoko, T. Watanabe, K. Takase, Y. Takano, *AIP Advances* **2018**, *8*, 101307.
12. H. Xiang, B. Xu, Y. Xia, J. Yin, Z. Liu, *RSC Advances* **2016**, *6*, 89901.
13. P. J. S. Foot, J. Suradi, P. A. Lee, *Materials Research Bulletin* **1980**, *15*, 189.
14. M. Piacentini, F. S. Khumalo, C. G. Olson, J. W. Anderegg, D. W. Lynch, *Chemical Physics* **1982**, *65*, 289.
15. P. A. Joy, S. Vasudevan, *Physical Review B* **1992**, *46*, 5425.
16. A. Ghosh, M. Palit, S. Maity, V. Dwij, S. Rana, S. Datta, *Physical Review B* **2021**, *103*, 064431.
17. J. min Zhang, Y. zhuang Nie, X. guang Wang, Q. lin Xia, G. hua Guo, *Journal of Magnetism and Magnetic Materials* **2021**, *525*, 167687.
18. D. Lançon, H. C. Walker, E. Ressouche, B. Ouladdiaf, K. C. Rule, G. J. McIntyre, T. J. Hicks, H. M. Rønnow, A. R. Wildes, *Physical Review B* **2016**, *94*, 214407.
19. Y. Zheng, X. X. Jiang, X. X. Xue, J. Dai, Y. Feng, *Physical Review B* **2019**, *100*, 174102.
20. J. U. Lee, S. Lee, J. H. Ryoo, S. Kang, T. Y. Kim, P. Kim, C. H. Park, J. G. Park, H. Cheong, *Nano Letters* **2016**, *16*, 7433.
21. K. S. Burch, D. Mandrus, J. G. Park, *Nature Publishing Group* **2018**, *563*, 47.
22. M. Gibertini, M. Koperski, A. F. Morpurgo, K. S. Novoselov, *Nature Publishing Group* **2019**, *14*, 408.
23. M. Ramos, F. Carrascoso, R. Frisenda, P. Gant, S. Mañas-Valero, D. L. Esteras, J. J. Baldoví, E. Coronado, A. Castellanos-Gomez, M. R. Calvo, *npj 2D Materials and Applications* **2021**, *5*, 1.
24. X. Han, P. Song, J. Xing, Z. Chen, D. Li, G. Xu, X. Zhao, F. Ma, D. Rong, Y. Shi, M. R. Islam, K. Liu, Y. Huang, *ACS Applied Materials and Interfaces* **2021**, *13*, 2836.
25. J. Chu, F. Wang, L. Yin, L. Lei, C. Yan, F. Wang, Y. Wen, Z. Wang, C. Jiang, L. Feng, J. Xiong, Y. Li, J. He, *Advanced Functional Materials* **2017**, *27*, 1701342.

26. R. N. Jenjeti, R. Kumar, M. P. Austeria, S. Sampath, *Scientific Reports* **2018**, *8*, 8586.
27. R. Kumar, R. N. Jenjeti, M. P. Austeria, S. Sampath, *Journal of Materials Chemistry C* **2019**, *7*, 324.
28. R. Kumar, R. N. Jenjeti, S. Sampath, *ACS Sensors* **2020**, *5*, 404.
29. Y. Fujii, A. Miura, N. C. Rosero-Navarro, M. Higuchi, K. Tadanaga, *Electrochimica Acta* **2017**, *241*, 370.
30. C. E. Byvik, B. T. Smith, B. Reichman, *Solar Energy Materials* **1982**, *7*, 213.
31. M. J. Coak, D. M. Jarvis, H. Hamidov, A. R. Wildes, J. A. M. Paddison, C. Liu, C. R. S. Haines, N. T. Dang, S. E. Kichanov, B. N. Savenko, S. Lee, M. Kratochvílová, S. Klotz, T. C. Hansen, D. P. Kozlenko, J. G. Park, S. S. Saxena, *Physical Review X* **2021**, *11*, 011024.
32. K. Z. Du, X. Z. Wang, Y. Liu, P. Hu, M. I. B. Utama, C. K. Gan, Q. Xiong, C. Kloc, *ACS Nano* **2016**, *10*, 1738.
33. Z. ur Rehman, Z. Muhammad, O. Adetunji Moses, W. Zhu, C. Wu, Q. He, M. Habib, L. Song, *Micromachines* **2018**, *9*, 292.
34. M. Nauman, D. H. Kiem, S. Lee, S. Son, J.-G. Park, W. Kang, M. J. Han, Y. J. Jo, *2D Materials* **2021**, *8*, 35011.
35. A. Hashemi, H. P. Komsa, M. Puska, A. v. Krasheninnikov, *Journal of Physical Chemistry C* **2017**, *121*, 27207.
36. McCreary, J. R. Simpson, T. T. Mai, R. D. McMichael, J. E. Douglas, N. Butch, C. Dennis, R. Valdés Aguilar, A. R. Hight Walker, *Physical Review B* **2020**, *101*, 064416.
37. A. R. Wildes, K. C. Rule, R. I. Bewley, M. Enderle, T. J. Hicks, *Journal of Physics Condensed Matter* **2012**, *24*, 416004.
38. G. Scandurra, J. Smulko, L. B. Kish, *Applied Sciences* **2020**, *10*, 5818.
39. J. Smulko, T. Chludziński, U. Çindemir, C. G. Granqvist, H. Wen, *Journal of Sensors* **2020**, *2020*, 1.
40. K. Geremew, S. Rumyantsev, F. Kargar, B. Debnath, A. Nosek, M. A. Bloodgood, M. Bockrath, T. T. Salguero, R. K. Lake, A. A. Balandin, *ACS Nano* **2019**, *13*, 7231.
41. G. Liu, S. Rumyantsev, M. A. Bloodgood, T. T. Salguero, A. A. Balandin, *Nano Letters* **2018**, *18*, 3630.
42. R. Salgado, A. Mohammadzadeh, F. Kargar, A. Geremew, C. Y. Huang, M. A. Bloodgood, S. Rumyantsev, T. T. Salguero, A. A. Balandin, *Applied Physics Express* **2019**, *12*, 037001.
43. S. Rumyantsev, M. Balinskiy, F. Kargar, A. Khitun, A. A. Balandin, *Applied Physics Letters* **2019**, *114*, 090601.
44. A. K. Geremew, S. Rumyantsev, M. A. Bloodgood, T. T. Salguero, A. A. Balandin, *Nanoscale* **2018**, *10*, 19749.
45. G. Liu, S. Rumyantsev, M. A. Bloodgood, T. T. Salguero, M. Shur, A. A. Balandin, *Nano Letters* **2017**, *17*, 377.
46. X. Wang, K. Du, Y. Y. F. Liu, P. Hu, J. Zhang, Q. Zhang, M. H. S. Owen, X. Lu, C. K. Gan, P. Sengupta, C. Kloc, Q. Xiong, *2D Materials* **2016**, *3*, 031009.
47. M. Scagliotti, M. Jouanne, M. Balkanski, G. Ouvrard, *Solid State Communications* **1985**, *54*, 291.
48. M. Scagliotti, M. Jouanne, M. Balkanski, G. Ouvrard, G. Benedek, *Physical Review B* **1987**, *35*, 7097.
49. K. Kim, J. U. Lee, H. Cheong, *Nanotechnology* **2019**, *30*, 452001.

50. S. Lee, K. Y. Choi, S. Lee, B. H. Park, J. G. Park, *APL Materials* **2016**, *4*, 086108.
51. G. Long, H. Henck, M. Gibertini, D. Dumcenco, Z. Wang, T. Taniguchi, K. Watanabe, E. Giannini, A. F. Morpurgo, *Nano Letters* **2020**, *20*, 2452.
52. Z. Ou, T. Wang, J. Tang, X. Zong, W. Wang, Q. Guo, Y. Xu, C. Zhu, L. Wang, W. Huang, H. Xu, *Advanced Optical Materials* **2020**, *8*, 2000201.
53. V. Grasso, F. Neri, S. Patanè, L. Silipigni, M. Piacentini, *Physical Review B* **1990**, *42*, 1690.
54. R. S. Haines, M. J. Coak, A. R. Wildes, G. I. Lampronti, C. Liu, P. Nahai-Williamson, H. Hamidov, D. Daisenberger, S. S. Saxena, *Physical Review Letters* **2018**, *121*, 266801.
55. A. Geremew, C. Qian, A. Abelson, S. Rumyantsev, F. Kargar, M. Law, A. A. Balandin, *Nanoscale* **2019**, *11*, 20171.
56. M. A. Stolyarov, G. Liu, S. L. Rumyantsev, M. Shur, A. A. Balandin, *Applied Physics Letters* **2015**, *107*, 023106.
57. A. K. Geremew, S. Rumyantsev, B. Debnath, R. K. Lake, A. A. Balandin, *Applied Physics Letters* **2020**, *116*, 163101
58. J. Renteria, R. Samnakay, S. L. Rumyantsev, C. Jiang, P. Goli, M. S. Shur, A. A. Balandin, *Applied Physics Letters* **2014**, *104*, 153104.
59. A. A. Balandin, *Nature Nanotechnology* **2013**, *8*, 549.
60. M. Z. Hossain, S. Rumyantsev, M. S. Shur, A. A. Balandin, *Applied Physics Letters* **2013**, *102*, 153512.
61. M. Z. Hossain, S. L. Rumyantsev, K. M. F. Shahil, D. Teweldebrhan, M. Shur, A. A. Balandin, *ACS Nano* **2011**, *5*, 2657.
62. V. Mitin, L. Regianni, L. Varani, Generation- recombination noise in semiconductors; Balandin, A. A.; Noise and Fluctuations Control in Electronic Devices; American Scientific Publishers Los Angeles, 2002; pp. 12 – 31.
63. J. Pavelka, J. Šikula, M. Tacano, M. Toita, *Radioengineering* **2011**, *20*, 194.
64. J. Pavelka, J. Šikula, M. Tacano, *Wseas Tranactions on Electronics* **2007**, *9*, 221.
65. S. Rumyantsev, G. Liu, M. S. Shur, R. A. Potyrailo, A. A. Balandin, *Nano Letters* **2012**, *12*, 2294.
66. J. Jaroszyński, D. Popović, T. M. Klapwijk, *Physical Review Letters* **2002**, *89*, 276401.
67. N. E. Israeloff, M. B. Weissman, G. J. Nieuwenhuys, J. Kosiorowska, *Physical Review Letters* **1989**, *63*, 794.
68. N. E. Israeloff, M. B. Weissman, G. A. Garfunkel, D. J. van Harlingen, J. H. Scofield, A. J. Lucero, *Physical Review Letters* **1988**, *60*, 152.
69. C. Reichhardt, C. J. O. Reichhardt, *Physical Review Letters* **2004**, *93*, 176405.
70. O. Cohen, Z. Ovadyahu, *Physical Review B* **1994**, *50*, 10442.
71. P. Dutta and P. M. Horn, *Reviews of Modern Physics*, **1981**, *53*, 497.